\documentclass[letter]{aa}

\usepackage{graphicx}
\usepackage{xcolor}
\usepackage{float}
\usepackage{txfonts}
\usepackage{subfig}
\usepackage{hyperref} 
\hypersetup{colorlinks=true,citecolor=blue}
\usepackage{lineno}

\usepackage[normalem]{ulem}

\newcommand{\G}{ \mathcal{G}}

\begin{document} 

   \title{Drifts of the sub-stellar points of the TRAPPIST-1 planets} 

   \author{Alexandre Revol\inst{1,2}
          \and
          Émeline Bolmont\inst{1,2}
          \and
          Mariana  Sastre\inst{3} 
          \and
          Gabriel Tobie\inst{4} 
          \and
          Anne-Sophie Libert\inst{5}
          \and
          Mathilde Kervazo\inst{4} 
          \and
          Sergi Blanco-Cuaresma\inst{6, 7} 
          }
   \institute{Observatoire de Genève, Université de Genève,
              51 Chemin Pegasi, 1290 Sauverny, Switzerland.
              \and
              Centre Vie dans l’Univers, Faculté des sciences, Université de Genève, Quai Ernest-Ansermet 30, 1211 Genève 4, Switzerland \\
              \email{alexandre.revol@unige.ch, emeline.bolmont@unige.ch}
              \and
              Kapteyn Astronomical Institute, University of Groningen, Netherlands
              \and
              Laboratoire de Planétologie et Géosciences, UMR-CNRS 6112, Nantes Université, 2 rue de la Houssinière, BP 92208, 44322 Nantes Cedex 3, France
              \and
              naXys, Department of Mathematics, University of Namur, 61 Rue de Bruxelles, 5000 Namur, Belgium
              \and
              Harvard-Smithsonian Center for Astrophysics, 60 Garden Street, Cambridge, MA 02138, USA
              \and
              Laboratoire de Recherche en Neuroimagerie, University Hospital (CHUV) , Lausanne, Switzerland
             }

   \date{Received 16/07/2024 / Accepted 17/09/2024}


    \abstract{
    Accurate modeling of tidal interactions is crucial for interpreting recent JWST observations of the thermal emissions of TRAPPIST-1~b and c and for characterizing the surface conditions and potential habitability of the other planets in the system. 
    Indeed, the rotation state of the planets, driven by tidal forces, significantly influences the heat redistribution regime.
    Due to their proximity to their host star and the estimated age of the system, the TRAPPIST-1 planets are commonly assumed to be in a synchronization state. 
    In this work, we present the recent implementation of the co-planar tidal torque and forces equations within the formalism of Kaula in the N-body code Posidonius.
    This enables us to explore the hypothesis of synchronization using a tidal model well suited to rocky planets.
    We studied the rotational state of each planet by taking into account their multi-layer internal structure computed with the code Burnman.
    Simulations show that the TRAPPIST-1 planets are not perfectly synchronized but oscillate around the synchronization state.
    Planet-planet interactions lead to strong variations on the mean motion and tides fail to keep the spin synchronized with respect to the mean motion.
    As a result, the sub-stellar point of each planet experiences short oscillations and long-timescale drifts that lead the planets to achieve a synodic day with periods varying from $55$~years to $290$~years depending on the planet.
   }


   \maketitle
%
\section{Introduction}\label{Sec:Introduction}

    First results of JWST observations of thermal emission of rocky exoplanets of the TRAPPIST-1 system \citep{Gillon2017} have been published, focusing on the characterization of the atmosphere of TRAPPIST-1~b \citep{greene_2023,Ih_2023} and TRAPPIST-1~c \citep{Zieba_2023_NothickCO2_T1c,Lincowski_2023_potential_atm_T1c}.
    To date, uncertainties persist about the presence and/or composition of the atmosphere of TRAPPIST-1~b and c, and further observations and data analysis will bring valuable insight to disentangle the state of the atmospheres.
    In this context, a good understanding of heat redistribution processes is crucial to accurately interpret the data for atmosphere's characterization.
    Given that the spin of the planets drives the heat redistribution regime and the tidal heating, and considering that the rotation states of the TRAPPIST-1 planets are driven by tides due to their close orbits to the star, accurately modeling these tidal interactions is essential for interpreting observations and inferring surface conditions.

    We present here recent developments made in the N-body code Posidonius\footnote{https://www.blancocuaresma.com/s/posidonius} \citep{Blanco-Cuaresma_2017,Bolmont2020b}.
    We implemented the formalism of \cite{Kaula_1964}, which allows us to study the spin evolution in a more relevant way for rocky planets compared to usual simple prescriptions of tides, such as constant time lag or constant phase lag model \citep[i.e. CTL and CPL;][]{Hut1981,Gold_Soter_1966,Goldreich1966}.
    Indeed, these models do not reproduce the correct behavior for highly viscous objects, especially for the evolution of their rotation \citep{Henning2009,Efroimsky2013}.
    In contrast, we use an Andrade rheology \citep{Andrade1910} and model the multi-layer internal structure of the TRAPPIST-1 planets using the code BurnMan \citep{Cottaar_2014,Myhill_2021}, assuming a rocky composition.
    We revisit the synchronized spin state of the TRAPPIST-1 planets and, in particular, the position of their sub-stellar point.
    For each planet, we compute the maximal theoretical drift due to the mean motion perturbation assuming a spin rate changing only due to the tidal lag.
    Section~\ref{Subsec:tidal_model} introduces the tidal formalism used (Kaula) and presents the equations implemented in the N-body code Posidonius.
    Internal structures used for the TRAPPIST-1 planets are described in Section~\ref{SubSec:Love_and_internal_structure}.
    Section~\ref{Sec:results} presents the results of the simulations, and shows the variation of the spin states around the synchronization and the drift of the sub-stellar point of each planet.
    
\section{Numerical setup}\label{Sec:setup}
    We use Posidonius \citep{Blanco-Cuaresma_2017,Bolmont2020b}, an N-body code that integrates the dynamical evolution of planetary systems.
    The code accounts for additional effects, such as tidal interactions, rotational flattening, general relativity, the evolution of the central star, and a migration prescription in a protoplanetary disk.
    In the original version of Posidonius, an equilibrium tide model \citep{Bolmont2015,Bolmont2020b} and a creep tide model \citep{Gomes2021}, were implemented.
    For this work, we only investigate the evolution of the system through planetary tide.
    The code does not take into account the triaxiality of the planets. 
    Then, we assume that the net restoring force from the permanent deformation is neglected.
    This assumption is discussed in Sec~\ref{Sec:conclusion}.
     For this study, we have implemented a new model of tides in (the N-body code)  Posidonius following the formalism of \cite{Kaula_1964}.
    The next section presents this model and the equations implemented in Posidonius.
    The code used in this study is available on GitHub\footnote{https://github.com/revolal/posidonius/tree/kaula\_v2} and it has been accepted to be merged into the next version of Posidonius.
    
    \subsection{Tidal model}\label{Subsec:tidal_model}

    We implemented the tidal forces and torques equations following the formalism of \cite{Kaula_1964}.
    This formalism is based on the expansion of the tidal perturbing gravitational potential in Fourier harmonic modes.
    Each mode of the expansion is associated with an excitation frequency and raises an associated tidal bulge.
    It makes this formalism general enough to encapsulate the frequency-dependent response of a body subjected to stress and strain \citep{efroimsky_Makarov_2014,boue_tidaltheory_2019}.
    According to \cite{Kaula_1964}, the tide-raising potential can be expressed in the Fourier domain in terms of Keplerian elements as 
    \begin{equation}
        \begin{split}
            W =& \sum_{l=2}^{\infty} \sum_{m=0}^l \sum_{p=0}^{l} \sum_{q\in \mathbb{Z}} W_{lmpq} (a, e, i,\omega_{lmpq}) ~, \\
        \end{split}
        \label{equ:Tidal_raising_potential_fourier_modes}
    \end{equation}
    The indexes $l,m,p,$ and $q$ are indexes of harmonic modes, $a, e, i,$ are the semi-major axis, eccentricity, and orbital inclination, respectively, and $\omega_{lmpq}$ the excitation modes, defined as
    \begin{equation}
        \omega_{lmpq} = (l-2p+q)\dot{\mathcal{M}} +(l-2p)\dot{\omega} +m(\dot{\Omega} -\dot{\theta})
    \end{equation} 
    with $\dot{\mathcal{M}}$ the mean anomaly derivative, $\dot{\omega}$ the argument of periastron derivative, $\dot{\Omega}$ the longitude of node derivative and $\dot{\omega}$ the spin rate of the planet \citep{efroimsky_Makarov_2014}.
    The tidal potential raised by a deformed planet can be determined from the tidal raising potential of Eq.~\ref{equ:Tidal_raising_potential_fourier_modes} as (we omit the dependence in $a, e, i,\omega_{lmpq}$ in the notation for simplicity)   
    \begin{equation}
        \begin{split}
            U_{lmpq} =& \Big(\frac{R_p}{r}\Big)^{l+1} \overline{k_l}(\omega_{lmpq})  W_{lmpq}\big|_{lag} ~, \\
        \end{split}
        \label{equ:Tidal_potential}
    \end{equation}
    with $R_p$ the radius of the planet, $r$ the radial distance between the star and the planet, $\overline{k_l}(\omega_{lmpq})$ the complex Love number described in Sec.~\ref{SubSec:Love_and_internal_structure}, and $U_{lmpq}\big|_{lag}$ the tidal raising potential of Eq.~\ref{equ:Tidal_raising_potential_fourier_modes}, where each phase is shifted with an angle $\epsilon_{lmpq}$ associated with an excitation mode $\omega_{lmpq}$. 
    The expression of the tidal force is then derived from the gravitational potential of Eq.~\ref{equ:Tidal_potential}.
    Neglecting the obliquity and orbital inclination, the tidal force can be expressed as
    \begin{equation}\label{eq:force_secul_coplan}
        \begin{split}
            \mathbf{F}&= -\frac{\G {M_\star}^2}{a^6}  \frac{R_p^{5}}{r} \sum_{q=-\infty}^{+\infty}  \Bigg[ \Big(\frac{3}{4}G_{21q}(e)^2 \Re(\overline{k_2}) +\frac{9}{4}G_{20q}(e)^2 \Re(\overline{k_2}) \Big)\mathbf{e_r} \\
             & +\frac{3}{2}G_{20q}(e)^2 \Im(\overline{k_2}) \mathbf{e_\varphi} \Bigg] ~, \\
        \end{split}
    \end{equation}
    with $\G$ the gravitational constant and $M_\star$ the mass of the host star.
    $G_{20q}(e)$ and $G_{20q}(e)$ are the eccentricity functions $G_{lpq}(e)$ with $l,m=2,0$ (see \citealt{Cayley_Tables-of-dvlp_1860} and appendix \ref{appendix:table_Kaula} for details).
    $\Re(\overline{k_2})$ and $\Im(\overline{k_2})$ are the real and imaginary parts of the complex Love number $\overline{k_l}$ at degree l=2 (we have omitted the frequency dependence in the formula for clarity), and $\mathbf{e_r}$ and $\mathbf{e_\varphi}$ are the radial and azimuthal unit vectors.
    The tidal torque applied to the planet is computed directly in the code from the tidal force with 
    \begin{equation}
        \mathbf{N} = \mathbf{r} \times \mathbf{F} ~,
        \label{eq:torque_cross_product}
    \end{equation}
    with $\mathbf{r}$ the vector $r~\mathbf{e_r}$.

    \subsection{Love numbers and internal structures of the Trappist-1 planets}\label{SubSec:Love_and_internal_structure}

    The complex Love number $\overline{k_l}(\omega_{lmpq})$ quantifies the response of a body submitted to tidal perturbations.
    The frequency dependence of the Love number associated with each planet is set as an input to the code, and interpolated in the code at each time-step for every excited frequency mode $\omega_{lmpq}$.
    We compute the frequency-dependent Love numbers using an internal structure model described in the following section.
    
        \subsubsection{Internal structure}\label{subsubsec:internal_structure}
\begin{figure*}[h]
            \centerline{\includegraphics[width=1.1\textwidth]{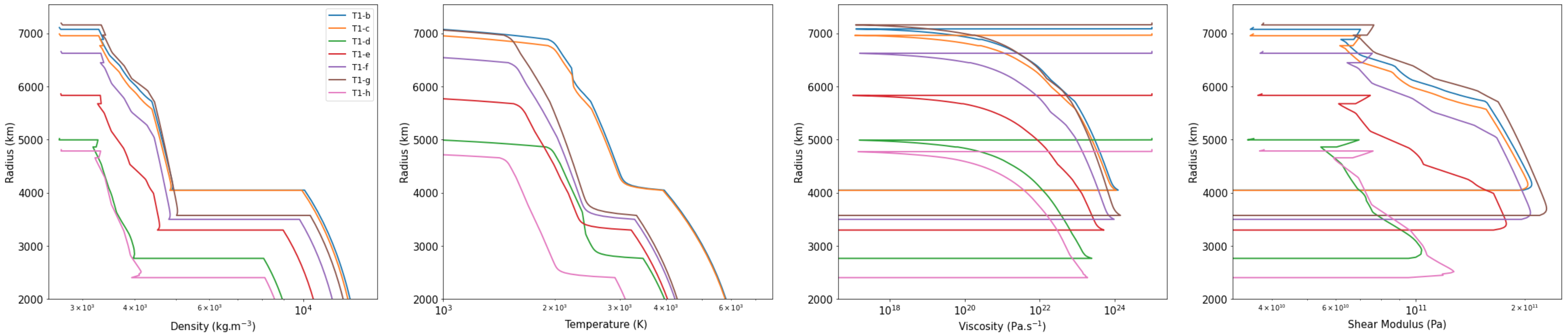}}
            \caption{Internal profiles of the TRAPPIST-1 planets computed with the BurnMan code.
            The figure shows, from left to right, the density profile,  the temperature profile, the viscosity profile, and the shear modulus profile.
            }
            \label{fig:t1_earth_internal}
        \end{figure*}

        The complex Love number can be computed for any internal structure, accounting for the density, shear modulus and viscosity profile.
        We computed its frequency dependence following the method described in \cite{Dumoulin2017}, \cite{Tobie_2005,Tobie2019}, and \cite{Bolmont2020a}.
        We considered a multi-layer model of planets assuming a compressible Andrade rheology \citep{Andrade1914}.
        The multi-layers internal structures of the TRAPPIST-1 planets are built with the BurnMan code\footnote{https://geodynamics.github.io/burnman/} \citep{Cottaar_2014,Myhill_2021}.
        The BurnMan code computes the density, temperature, gravity, and pressure profiles from the planet's surface to its center.
        The viscosity profile in the mantle is estimated from the temperature profile, computed with the Rayleigh number with the value at the bottom of the mantle of the Earth ($R_a = 10^6$).
        The viscosity of the lithosphere is set at $10^{25}$~Pa.s.
        Internal structures are computed from possible compositions and core sizes compatible with the mass and radius estimations of TRAPPIST-1 planets from \cite{Agol2021}.
        The masses and radii can be reproduced with a variety of internal structures and compositions.
        Thus, for each planet, we assume a silicate mantle with pyrolitic composition, and a liquid metallic core with an amount of iron and silicate of 90\% and 2\% respectively, which corresponds to the Earth \citep{hirose_light_2021}. 
        With the composition fixed, we vary the relative radius of the core of each planet from 50 to 60\% in order to match their total mass and radius.
        The masses and radius in Table~\ref{tab:trappist1_earth_internal}.
        Figure~\ref{fig:t1_earth_internal} shows the temperature, density, viscosity, and rigidity profiles for each TRAPPIST-1 planet computed with the BurnMan code.
    
        \begin{table*}[h!]
            \centering
            \caption{TRAPPIST-1 system parameters used as initial conditions for the simulations. The orbital parameters, i.e. $a$ the semi-major axis, $e$ the eccentricity, $\varpi$ the longitude of ascending node, and $\lambda$ the mean longitude, are taken from \cite{Agol2021}}
            \label{tab:trappist1_earth_internal}
            \begin{tabular}{|c|c|c|c|c|c|c|c|}
            \hline
                Star & Mass M$_\odot$ & Radius R$_\odot$  & - & - & - & - \\
                \hline
                T1- A& 0.0898 M$_\odot$ & 0.1192 R$_\odot$  & - & - & - & - \\
                \hline
                Planets & Mass M$_\oplus$ & Radius R$_\oplus$  & $a$ ($10^{-2}$~AU) & $e$ ($10^{-3}$) & $\varpi$ (rad) & $\lambda$ (rad)  \\
                \hline
                T1-b &1.374 &1.116 &1.154 &6.694 &3.9138 &6.2807 \\
                T1-c &1.308 &1.097 &1.580 &1.958 &3.9568 &3.2752 \\
                T1-d &0.388 &0.788 &2.227 &7.221 &3.6462 &0.2445 \\
                T1-e &0.692 &0.920 &2.927 &5.215 &0.9017 &1.4578 \\
                T1-f &1.039 &1.045 &3.851 &9.502 &3.1105 &1.0048 \\
                T1-g &1.129 &1.129 &4.684 &3.263 &5.8092 &1.4551 \\
                T1-h &0.755 &0.755 &6.192 &4.403 &3.2359 &5.0763 \\
    	        \hline
    	   \end{tabular}
        \end{table*}

    \subsubsection{Love numbers}\label{subsubsec:Love_numbers}

\begin{figure}[h]
            \centering
            \includegraphics[width=0.45\textwidth]{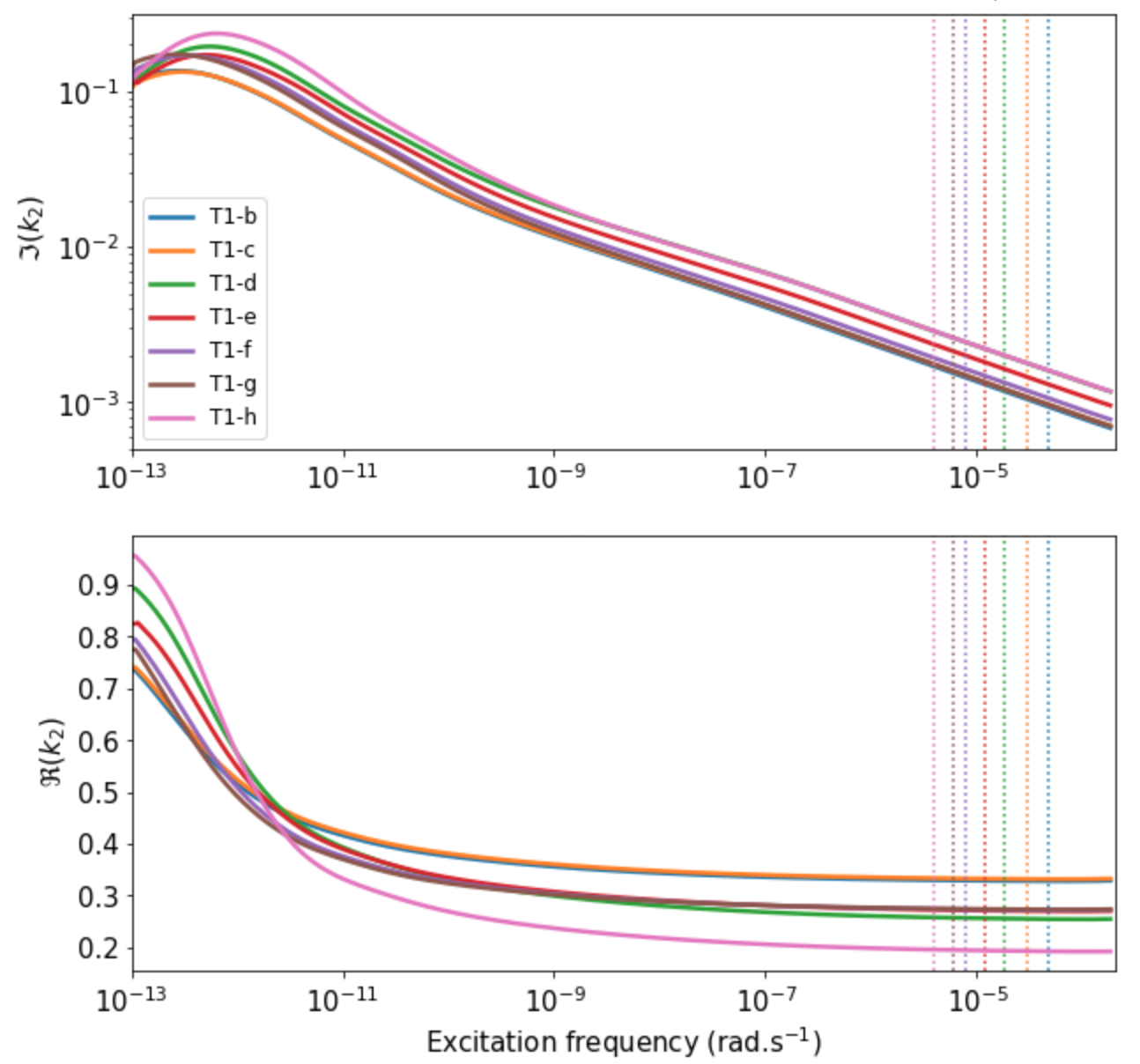}
            \caption{Frequency dependence of the imaginary part $\Im(k_2)$ (top panel) and real part $\Re(k_2)$ (bottom panel) of the Love number of the TRAPPIST-1 planets assuming Earth-like compositions, computed from the profiles presented in Sec.~\ref{subsubsec:internal_structure} and Fig~\ref{fig:t1_earth_internal}.
            The vertical dotted lines correspond to each planet's principal excitation frequency (i.e. the mean motion $n$) if considered as perfectly synchronous.}
            \label{fig:love_number_trappist_earth}
        \end{figure}

        Love numbers are computed following the method of \cite{Tobie_2005,Tobie2019} with the code LPG-Tide, which computes them from the radial profile of the density, viscosity, shear modulus, and seismic velocities.
        We use an Andrade rheology \citep{Andrade1910}, which defines the complex compliance $\Bar{J}$ as \citep{Castillo-Rogez2011, Efroimsky2012a}
        \begin{equation}
            \bar J  = J +\beta(i \omega)^{-\alpha}\Gamma(1+\alpha) -\frac{i}{\eta \omega} ~,
            \label{eq:andrade_complex_compliance}
        \end{equation}
        with $\eta$ the shear viscosity, $\omega$ the excitation frequency, $\beta$ and $\alpha$ two empirical parameters, and $\Gamma$ the Gamma function.
        Figure~\ref{fig:love_number_trappist_earth} shows the Love numbers frequency dependences associated with the TRAPPIST-1 profiles presented in Sec.~\ref{subsubsec:internal_structure}.
        The top panel of the Fig.~\ref{fig:love_number_trappist_earth} represents the frequency dependence of the imaginary part of the Love number, thus the dissipation within each body \citep{Tobie_2005,Dumoulin2017,Bolmont2020a}.
    
\section{TRAPPIST-1 rotational states}\label{Sec:results}

    We performed N-body simulations of the TRAPPIST-1 system.
    The rotational state is assumed to be synchronous at the beginning of the simulation.
    The orbital inclinations and obliquities of the planets are assumed to be zero, considering the system's high coplanar configuration and high tidal damping.
    The initial conditions are taken from \cite{Agol2021} and listed in Table~\ref{tab:trappist1_earth_internal}.
    The following section presents the evolution of the rotation state of each planet, their deviation from the synchronization state, and the variation of the position of the sub-stellar points.

    \subsection{Non-synchronization}\label{Subsec:non-synchro}
        
\begin{figure}[h]
            \centering
            \includegraphics[width=0.5\textwidth]{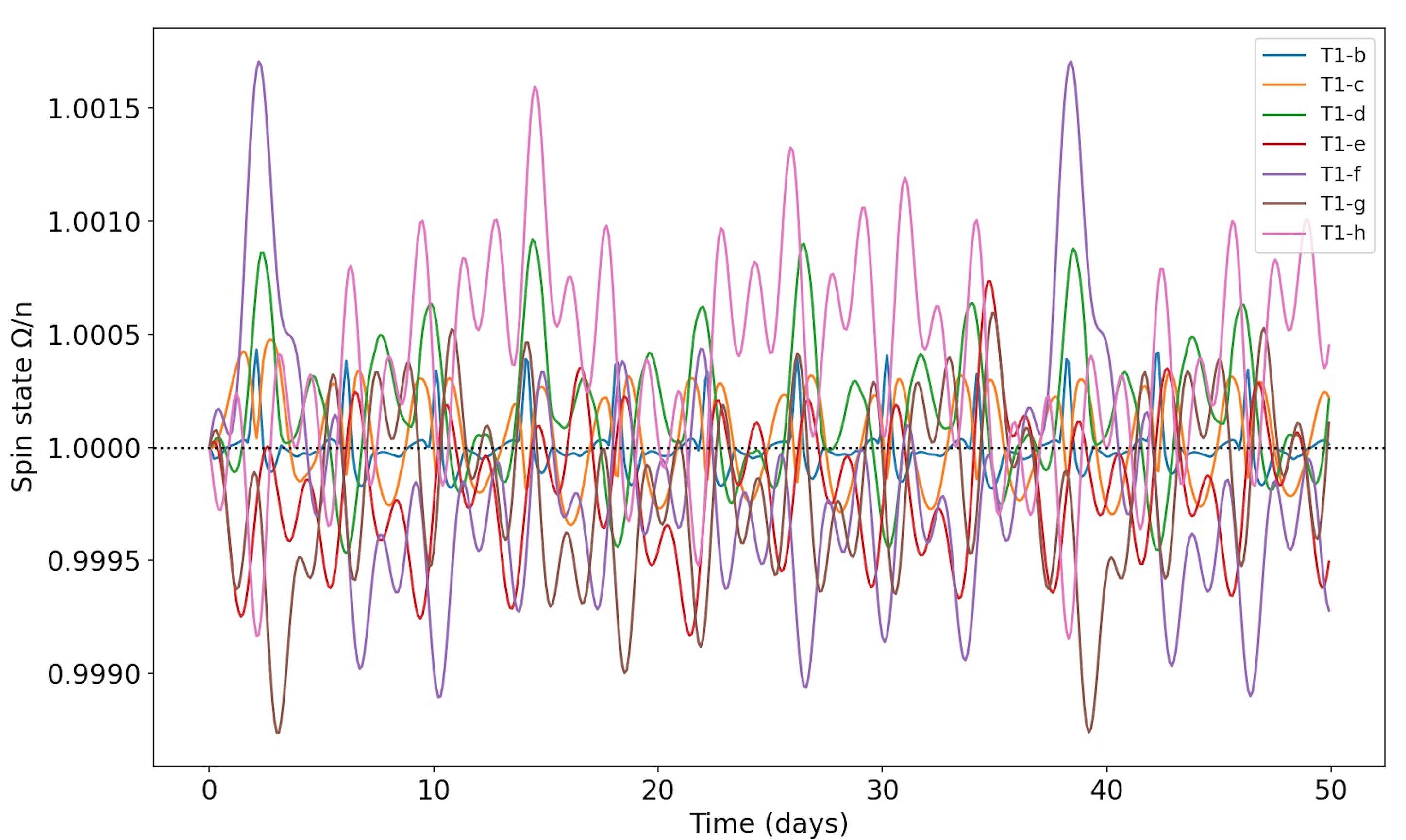}
            \caption{Evolution of the rotation state of the TRAPPIST-1 planets in terms of $\Omega/n$ (the spin rate and the mean motion, respectively) around the synchronization state ($\Omega/n=1$).}
            \label{fig:spin_variation_synchro}
        \end{figure}

\begin{figure}[h]
            \centering
            \includegraphics[width=0.45\textwidth]{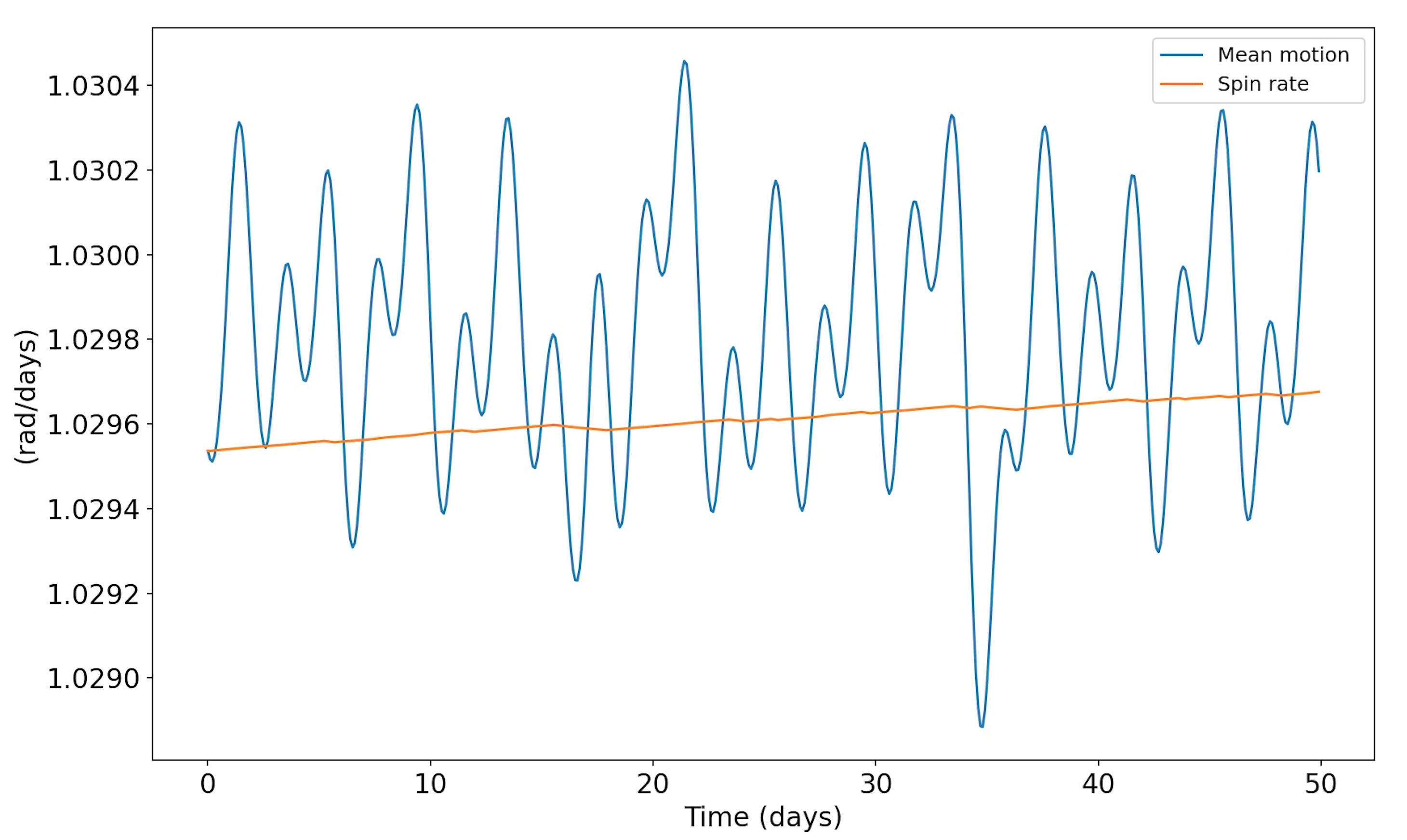}
            \caption{Spin rate and mean motion of the planet e over 50~days.}
            \label{fig:t1e_spin_meamotion}
        \end{figure}

        Figure~\ref{fig:spin_variation_synchro} shows the variation of the rotation state, in terms of $\Omega/n$ (the spin rate and the mean motion, respectively), around the synchronization state for a period of 50~days.
        The simulation shows that none of the planets is perfectly synchronized, but oscillates around the synchronization state, i.e. $\Omega/n=1$, with different amplitudes and timescales.
        These variations are due to large mean motion variations forced by planet-planet interactions.
        As shown in Fig.~\ref{fig:t1e_spin_meamotion} for the planet e, the mean motion varies with such high amplitudes and short timescales, that it prevents the synchronization of the spin.

    \subsection{Sub-stellar point drift}\label{Subsec:sub-stellar}

\begin{figure}[h]
            \centering
            \includegraphics[width=0.5\textwidth]{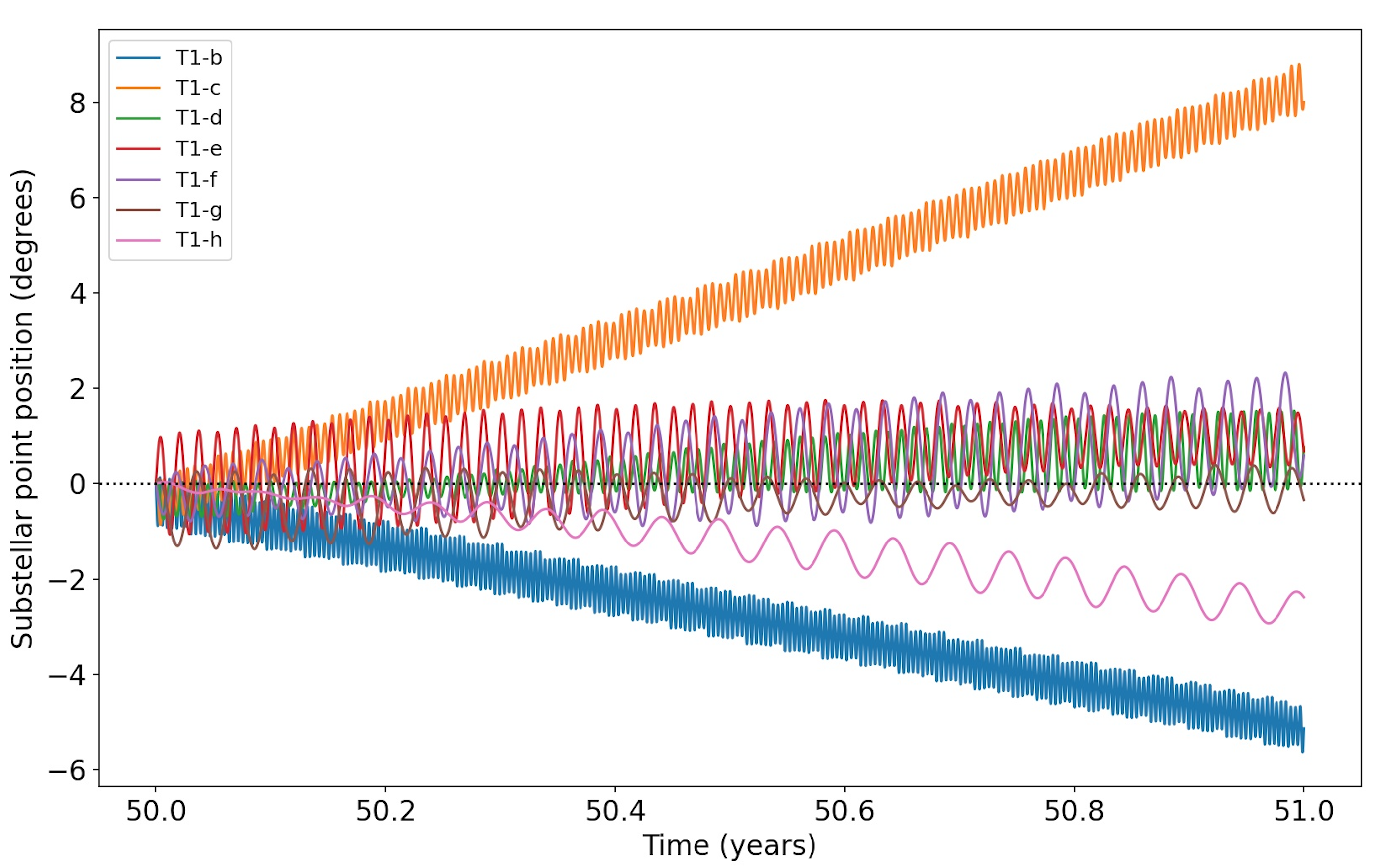}
            \caption{Short-term librations of sub-stellar points of the TRAPPIST-1 planets, in degrees, for a period of 1~year.
            Positive angles correspond to eastward, while negative angles correspond to westward drifts. 
            An eastward, respectively westward, rotation corresponds to prograde rotation, respectively retrograde rotation, of the planet regarding its orbit.}
            \label{fig:sub-stellar-point_zoom}
        \end{figure}

\begin{figure}[h]
            \centering
            \includegraphics[width=0.5\textwidth]{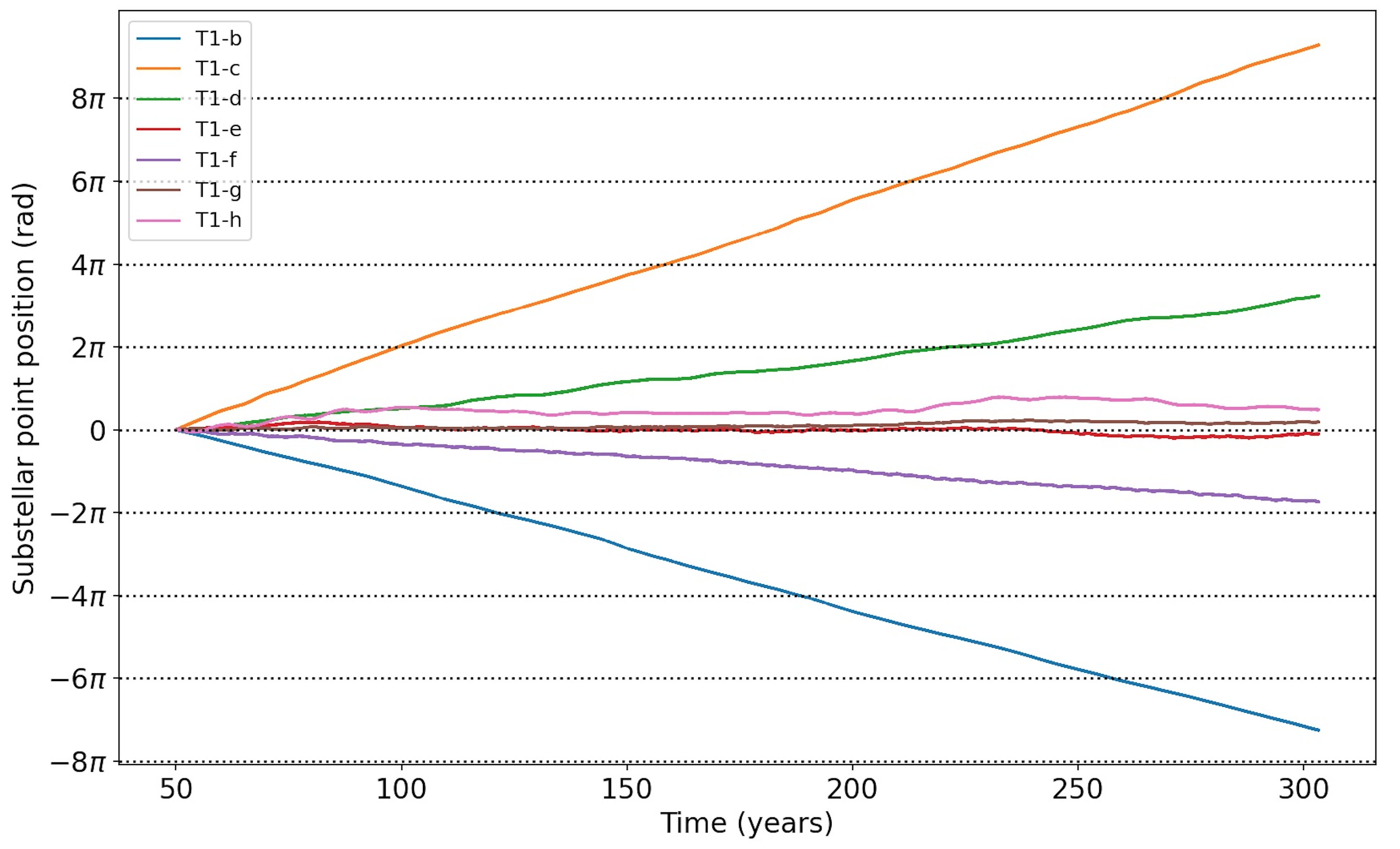}
            \caption{Long-term variation of sub-stellar points of the TRAPPIST-1 planets for a period of 250~years.
            Horizontal dotted lines represent multiples of $2\pi$ rotation.}
            \label{fig:sub-stellar-point}
        \end{figure}

        We computed the sub-stellar position of each planet by subtracting {$\nu-\theta$}, with $\theta$ the rotation angle, computed from the integration of the spin rate $\dot{\theta}$, and $\nu$ the position angle of the star in the sky of the planet (the true anomaly).
        Sub-stellar points are computed to be zero at the beginning after 50~years of simulation.
        Variations of the spin states around the synchronization lead to short- and long-term variations of sub-stellar points.
        Figure~\ref{fig:sub-stellar-point_zoom} shows short-term variations, which correspond to short variations of the mean motion, as shown in Fig.~\ref{fig:t1e_spin_meamotion}.
        Amplitudes of low librations vary between 1 and 2 degrees depending on the planet, with a timescale that varies from about 1.5~days for the planet~b to 18 days for the planet~h, which corresponds approximately to the orbital period of each planet.
        Figure~\ref{fig:sub-stellar-point} shows long-term drifts of the sub-stellar point of each planet for a period of 250~years.
        The time needed for the sub-stellar points to perform a complete rotation (i.e., the duration of the synodic day) varies between 55~years and -290~years, depending on the planet, and is summarized in the Table~\ref{tab:length_of_day}.
        The estimations of planets T1-e, g and h (in gray) should be taken with caution, as they appear to be strongly tidally locked and have not completed a single synodic rotation in 250 yrs.
        Positive drift corresponds to a retrograde rotation with respect to its orbit, while a negative drift corresponds to a prograde rotation, respectively.
        We must highlight that the formalism used does not take into account the triaxiality of the planets, then, considering the absence of the restoring force from the permanent deformation, these estimations correspond to the maximum theoretical drift due to the mean motion variations.
        Accounting for the triaxiality should lead to a stronger tidal torque \citep{VanHoolst_2013} and thus, reducing the drifts rate.
        The triaxiality, which is expected to be particularly pronounced for this close-in planet due to strong permanent tidal forces, should further limit the drift.
        Interestingly, the two inner planets have the strongest drifts in the system, despite their closer distance from the star, and are moving in the opposite way.
        We found that it is link to the position of the planets in the resonance chain that lead to strong planet-planet interactions.
        Planet~b also shows numerical artifacts which make the computation of the sub-stellar point challenging for a symplectic integrator.
        The combination of the very short orbital period and the shape of the Love number makes the dissipation located in the low-frequency regime on the dissipation spectrum (see top panel of Fig.~\ref{fig:love_number_trappist_earth}).
        
        
        \begin{table}[h!]
            \centering
            \caption{Estimation of the length of the day of the TRAPPIST-1 planet according to the rotation period of their sub-stellar point. Numbers in gray are estimations from linear fits and should be taken with caution.}
            \label{tab:length_of_day}
            \begin{tabular}{|c|c|}
            \hline
                Planet & Synodic days (years) \\
                \hline
                T1-b & -69   \\
                T1-c & 55    \\
                T1-d & 160   \\
                T1-e & {\color{gray}-1997} \\
                T1-f & -290  \\
                T1-g & {\color{gray}2400} \\
                T1-h & {\color{gray}1103} \\
                \hline
           \end{tabular}
        \end{table}

\section{Conclusion}\label{Sec:conclusion}

    This letter aims to present recent developments made in the code Posidonius \citep{Blanco-Cuaresma_2017, Bolmont2020a}, i.e. the implementation of the formalism of Kaula, that accounts for a realistic frequency dependence of the tidal response \citep{Kaula_1964}.
    We applied our work to the TRAPPIST-1 system and studied for the first time the variation of the spin of each planet around the synchronization state accounting for the forcing of the tidal bulge only assuming a realistic tidal model.
    We model the tidal response of each planet with the rheology of \cite{Andrade1910} following the method of \cite{Bolmont2020a} with a multi-layered internal structure computed with the internal structure code BurnMan \citep{Cottaar_2014,Myhill_2021}.

    We found that, despite the strong tidal interactions in the system, the spin of each planet is not perfectly synchronized (i.e. they don't have a spin rate strictly equal to the mean motion).
    Planet-planet interactions significantly perturb the mean motion of planets in compact configurations that prevent spin rates from remaining synchronous, causing them to oscillate around the synchronization state.
    As a consequence of these oscillations, planets experience variations of the position of their sub-stellar point, leading to a slow and continuous drift.
    Continuous drifts lead the sub-stellar points to perform complete rotations, which lead to day-night cycles, on retrograde rotation, with synodic day periods varying from 55 to 290 of years depending on the planet.
    
    We must highlight that the triaxiality of each planet is not taken into account in this study.
    It is still uncertain how the long-term shape of the planet should evolve with respect to the spin-orbit evolution. 
    On the one hand, if the spin-orbit evolves on a timescale much smaller than the interior relaxation timescale, the planet would not have time to develop a permanent long-axis deformation in the star direction and the planet should remain axisymetric. 
    On the other hand, if the spin-orbit evolves on much longer timescale than the relaxation timescale, a permanent long-axis would develop and counteract any planet drift. 
    Even in this condition, a drift is still possible, the drift rate will be controlled by the rate at which the planet shape can readjust as the spin-orbit change.  
    As we neglect the distorted shape of the planets and their long-term evolution, the estimations we obtained here should be considered as the maximum synodic days from the spin oscillations around the synchronization state.
    In particular, \cite{Correia2019} evaluated the damping of the triaxiality due to mean motion variations, but assumed a fixed permanent deformation (assuming that the triaxiality of the planet has two components, the permanent deformation due to the intrinsic mass repartition in the planet and the tidally induced deformation, see also \citealt{Delisle_2017} appendix~D).

    Non-synchronous spin states of TRAPPIST-1 planets have already been suggested a first time by \cite{Makarov2018}, as spin-orbit resonances and/or pseudo-synchronization from the eccentricities of the system.
    However, our findings involve not only eccentricities but the overall non-Keplerian shapes of the orbits, due to planet-planet interactions. 
    These interactions have been studied by \cite{vinson_chaotic_2019}, \cite{shakespeare_day_2023} and \cite{Chen2023_SporadicSpinVar}, but their results differ from our findings on several points.
    
    In particular, \cite{vinson_chaotic_2019} and \cite{shakespeare_day_2023} identified several regimes of rotation coming from the pendulum effect from the triaxiality of planets.
    But firstly, they did not compute the spin rate in a full consistent way with their N-body simulation.
    Secondly, the tidal model used was the common constant time lag model \citep[CTL,][]{Hut1981,Eggleton_1998}, which is not well-suited to study the rotation of rocky planets submitted to tides \citep{Makarov2013}.
    
    This study is the first to analyze the spin state of rocky planets that account for their internal structure. 
    The drifts of sub-stellar points depend on the competition between mean motion variation and tidal damping timescales, which are influenced by the strength of the tides and thus, by the internal structure, and the potential presence of a surface magma or water ocean. 
    Future work will further investigate the trend of the substellar drifts and provide a detailed analysis of the impact of tidal models used, the rheological properties, and internal structures of TRAPPIST-1 planets.
    In this study, we assume TRAPPIST-1 planets have a metallic core and a silicate mantle, particularly relevant for the inner planets (b, c, d). 
    However, the low bulk density of the outer planets (e, f, g) suggests they have significant volatile content and may also be compatible with the presence of water or ice \citep[e.g.][]{Unterborn2018,Boldog2024}.  
    Exploring a broader range of internal structures is beyond this paper's scope.
    These insights are crucial for understanding the physical and thermal processes of these exoplanets, as well as in the interpretation of observational data and assessing their potential habitability.

 
\begin{acknowledgements}
    AR and EB acknowledge the financial support of the SNSF (grant number: 200021\_197176 and 200020\_215760).
    The authors thank the referee, Michael Efroimsky, who provided good criticisms that help to improve our work.
    This work has been carried out within the framework of the NCCR PlanetS supported by the Swiss National Science Foundation under grants 51NF40\_182901 and 51NF40\_205606.
    The computations were performed at University of Geneva on the Baobab and Yggdrasil clusters.
    This research has made use of NASA's Astrophysics Data System.
\end{acknowledgements}

%
%

\bibliographystyle{aa}
\bibliography{Biblio}

\begin{appendix} 
       
\section{Table of the inclination eccentricity from the Fourier development by Kaula, Cayley etc}\label{appendix:table_Kaula}

    The eccentricity polynomials $G_{lpq}(e)$ are elliptic expansions given by Eqs.~$23$ and $24$ of \cite{Kaula_1961} and can be computed from the Hansen functions $\mathbf{X}_{l-2p+q}^{-(l-1),(l-2p)}$ \citep{Cayley_Tables-of-dvlp_1860,Tisserand1889} and presented in the Table~\ref{tab:G_eccentricite}.
    
    We considered eccentricities up to $0.3$, which allowed us to consider the eccentricity expansions up to order $7$ \citep[see tables][]{Cayley_Tables-of-dvlp_1860}.
    As the tidal interactions are computed at the quadupolar order $l=2$, the index $p$ is constrained between $0$ and $2$ and $q$ between $-7$ and $7$.
    The elements $G_{22q}(e)$ can be computed with the symmetry rule $G_{2,2,q}(e) = G_{2,0,-q}(e) $.
    \begin{table}[h]
        \caption{Eccentricity polynomials $G_{lpq}(e)$ from \citet[]{Cayley_Tables-of-dvlp_1860}, up to order $7$ in eccentricity.}
        \label{tab:G_eccentricite}
         \centering
         \begin{tabular}{| c | c | c || l |}
            \hline
               l & p & q & $G_{lpq}(e)$ \\
       \hline
       2 & 0 & -7 & $ \frac{15625}{129024}e^7 $  \\
       2 & 0 & -6 & $ \frac{4}{45}e^6 $ \\
       2 & 0 & -5 & $ \frac{81}{1280}e^5 +\frac{81}{2048}e^7  $ \\
       2 & 0 & -4 & $ \frac{1}{24}e^4 +\frac{7}{240}e^6 $ \\
       2 & 0 & -3 & $ \frac{1}{48}e^3 +\frac{11}{768}e^5 +\frac{313}{30720}e^7 $    \\
       2 & 0 & -2 & $0 $  \\
       2 & 0 & -1 & $-\frac{1}{2}e +\frac{1}{16}e^3 -\frac{5}{384}e^5 -\frac{143}{18432}e^7$     \\
       2 & 0 & 0 & $ 1 -\frac{5}{2}e^2 +\frac{13}{16}e^4 -\frac{35}{288}e^6 $   \\
       2 & 0 & 1 & $ \frac{7}{2}e -\frac{123}{16}e^3 +\frac{489}{128}e^5 -\frac{1763}{2048}e^7 $   \\
       2 & 0 & 2 & $ \frac{17}{2}e^2 -\frac{115}{16}e^4 +\frac{601}{48}e^6$   \\
       2 & 0 & 3 & $ \frac{845}{48} e^3 -\frac{32525}{768} e^5 +\frac{208225}{6144} e^7 $    \\
       2 & 0 & 4 & $ \frac{533}{16} e^4 -\frac{13827}{160} e^6 $         \\
       2 & 0 & 5 & $ \frac{228347}{3840} e^5 -\frac{3071075}{18432} e^7 $   \\
       2 & 0 & 6 & $ \frac{73369}{720} e^6 $         \\
       2 & 0 & 7 & $ \frac{12144273}{71680} e^7 $    \\
       \hline
       2 & 1 & -7 & $G_{217}(e)$   \\
       2 & 1 & -6 & $G_{216}(e)$   \\
       2 & 1 & -5 & $G_{215}(e)$   \\
       2 & 1 & -4 & $G_{214}(e)$   \\
       2 & 1 & -3 & $G_{213}(e)$   \\
       2 & 1 & -2 & $G_{212}(e)$   \\
       2 & 1 & -1 & $G_{211}(e)$    \\
       2 & 1 & 0 & $ (1 -e^2)^{-3/2} \simeq 1 +\frac{3}{2}e^2 +\frac{15}{8}e^4 +\frac{35}{16}e^6 +\mathcal{O}(e^9)$    \\
       2 & 1 & 1 & $ \frac{3}{2}e +\frac{27}{16}e^3 +\frac{261}{128}e^5 +\frac{14309}{6144}e^7$    \\
       2 & 1 & 2 & $ \frac{9}{4}e^2 +\frac{7}{4}e^4 +\frac{141}{64}e^6$    \\
       2 & 1 & 3 & $\frac{53}{16}e^3 +\frac{393}{256}e^5 +\frac{24753}{10240}e^7 $  \\
       2 & 1 & 4 & $\frac{77}{16}e^4 +\frac{129}{160}e^6 $  \\
       2 & 1 & 5 & $\frac{1773}{256}e^5 -\frac{4987}{6144}e^7 $  \\
       2 & 1 & 6 & $\frac{3167}{320}e^6 $   \\
       2 & 1 & 7 & $\frac{432091}{30720}e^7 $   \\
       \hline  
         \end{tabular}
     \end{table}
     
     In practice, the order of the summation over $q$ is taken in function of the value of the eccentricity. The maximum order considered is listed in Table \ref{tab:ordrer_ecc}.
     \begin{table}[h]
         \centering
         \begin{tabular}{|c|c|c|c|c|c|c|c|c|}
         \hline
              e & 0 & <0.05 & <0.1 & <0.15 & <0.2 & <0.25 & <0.3 & >0.3 \\
              \hline
              |q|$_{\text{max}}$ & 0 & 1 & 2 & 3 & 4 & 5 & 6 & 7 \\
              \hline
         \end{tabular}
         \caption{Maximum order of the summation over $q$ in function of the eccentricity.}
         \label{tab:ordrer_ecc}
     \end{table}

\end{appendix}

\end{document}